\documentclass[aps,prl,twocolumn,superscriptaddress,floatfix,showpacs,groupedaddress]{revtex4-1}
\usepackage{graphicx}
\usepackage{dcolumn}
\usepackage{epstopdf}
\usepackage{hyperref}
\usepackage{bm}
\usepackage{color}
\usepackage{amsfonts}
\usepackage{amssymb}
\usepackage{amsmath}
\usepackage{txfonts}
\usepackage{lipsum}
\usepackage{wasysym}
\usepackage{verbatim}

\bibpunct{[}{]}{,}{n}{}{}

\newcommand{\qv}{\mathbf{q}}
\newcommand{\kv}{\mathbf{k}}

\begin{document}

\title{Excitonic Instability and Pseudogap Formation in Nodal Line Semimetal ZrSiS}

\author{A.~N. Rudenko$^{1,2,4}$}
\email[]{a.rudenko@science.ru.nl}
\author{E.~A. Stepanov$^{2,4}$}
\author{A.~I. Lichtenstein$^{3,4}$}
\author{M.~I. Katsnelson$^{2,4}$}

\affiliation{\mbox{$^1$School of Physics and Technology, Wuhan University, Wuhan 430072, China}\\
$^2$\mbox{Institute for Molecules and Materials, Radboud University, Heijendaalseweg 135, 6525 AJ Nijmegen, The Netherlands}\\
$^3$\mbox{Institute for Theoretical Physics, University of Hamburg, Jungiusstrasse 9, D-20355 Hamburg, Germany}\\
$^4$\mbox{Theoretical Physics and Applied Mathematics Department, Ural Federal University, Mira Street 19, 620002 Ekaterinburg, Russia}}

\date{\today}

\begin{abstract}
Electron correlation effects are studied in \mbox{ZrSiS} using a combination of first-principles and model approaches. We show that basic electronic properties of \mbox{ZrSiS} can be described within a
two-dimensional lattice model of two nested square lattices. A high degree of electron-hole symmetry characteristic for ZrSiS is one of the key features of this model.
Having determined model parameters from first-principles calculations, we then explicitly take electron-electron interactions into account and show that, at moderately low temperatures, \mbox{ZrSiS} exhibits excitonic instability, leading to the formation of a pseudogap in the electronic spectrum. The results can be understood in terms of Coulomb-interaction-assisted pairing of electrons and holes reminiscent of that of an excitonic insulator. Our finding allows us to provide a physical interpretation of the unusual mass enhancement of charge carriers in \mbox{ZrSiS} recently observed experimentally.
\end{abstract}

\maketitle
\mbox{ZrSiS} belongs to a class of three-dimensional materials that feature a nodal line in the electronic band structure, resulting from linearly dispersed bands forming cones reminiscent of the Dirac
cones \cite{Neupane,Schoop,Hu}. Recently, \mbox{ZrSiS} has been extensively studied experimentally, demonstrating a rich variety of remarkable properties \cite{Wang,Ali,Singha,Matusiak,Lodge,Butler}. Among these are a strongly nested Fermi surface \cite{Neupane,Schoop}, a large $g$ factor \cite{Hu}, topologically nontrivial states \cite{Wang,Singha,Ali}, and unconventional mass enhancement of quasiparticles
near the nodal line, as indicated by quantum oscillations measured in high magnetic fields \cite{Pezzini}. The latter finding suggests the importance of electron correlation effects in the properties of \mbox{ZrSiS}, which may contribute to the realization of exotic quantum states in this material.

In this Letter, we study many-body effects in \mbox{ZrSiS}. We begin with constructing a tractable lattice model on the basis of density functional theory (DFT) calculations. We then consider a two-orbital Hubbard Hamiltonian within the ladder approximation in the form suggested in Refs.~\onlinecite{FLEX1,FLEX2}. We find that at moderately low temperatures \mbox{ZrSiS} enters in the regime of exciton instability, leading to the formation of a pseudogap in the electronic spectrum.

\begin{figure}[tpb]
\includegraphics[width=0.95\linewidth]{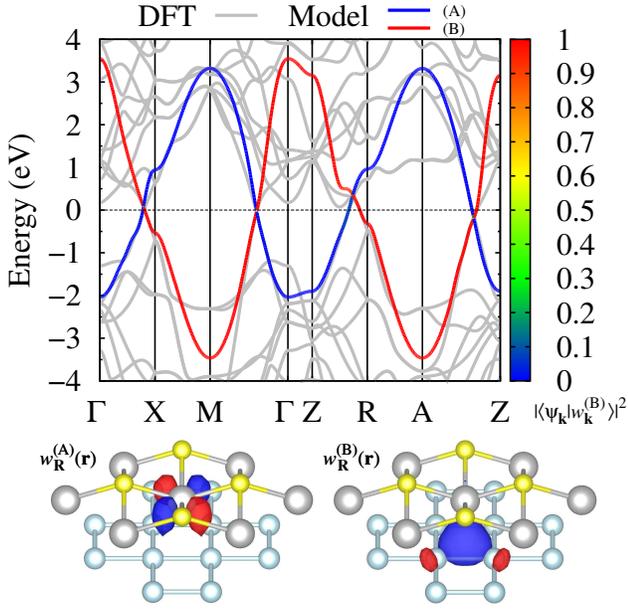}
\caption{(Top) Band structure of \mbox{ZrSiS} calculated by projecting the DFT Hamiltonian onto WFs obtained within the scheme of maximal localization. Color indicates projection of Bloch states on WFs forming sublattices $A$ and $B$. Original DFT structure is shown for reference. Notation of high-symmetry points of the Brillouin zone is the same as in Ref.~\onlinecite{Schoop}. Horizontal dashed line corresponds to the Fermi energy. (Bottom) Real-space representation of WFs being the basis of the TB Hamiltonian used in this Letter to describe many-body effects in ZrSiS. Gray, blue, and yellow balls depict Zr, Si, and S atoms, respectively.}
\label{bands}
\end{figure}

{\it Model.}---We first calculate the electronic structure of \mbox{ZrSiS} using DFT. Specifically,
we used generalized gradient approximation \cite{PBE} in combination with
the projected augmented-wave method \cite{PAW} as implemented in the {\sc vasp} code \cite{Kresse1,Kresse2}. Kinetic energy cutoff was
set to 200 eV and the Brillouin zone was sampled by a ($16\times16\times8$) zone-centered {\bf k}-point mesh.
Experimental lattice parameters that were used were obtained from single-crystal diffraction studies \cite{Schoop}.
The resulting band structure is shown in Fig.~\ref{bands}, which is in full agreement with other studies, including Ref.~\onlinecite{Schoop}.
In the vicinity of the Fermi energy, one can see the characteristic feature of the \mbox{ZrSiS} band structure that is conical intersections of bands forming the line of crossing points (nodes).

To make the description of the \mbox{ZrSiS} single-particle electronic structure more tractable, we aim at the construction of a low-energy tight-binding (TB) model.
To this end, we exploit the formalism of maximally localized Wannier functions (WFs) \cite{Marzari,Souza,MLWF}, which allows us to find a unitary transformation that projects the DFT eigenstates
onto the set of Bloch functions that are supposed to describe bands in the relevant energy region.
Having done this, one can construct the TB Hamiltonian defined in the optimal WF basis. A minimal model that would describe the nodal line structure of \mbox{ZrSiS} requires two orbitals.
Taking this into account, we utilize the criterion of maximal localization in order to find two relevant WFs for \mbox{ZrSiS} \cite{wannier_code}, whose resulting real-space representation is shown in Fig.~\ref{bands}.
One orbital $w_{\bf R}^{(A)}({\bf r})$ is localized on the Zr atom and resembles a cubic harmonic of $d_{xy}$ symmetry, whereas the other orbital $w_{\bf R}^{(B)}({\bf r})$ is more delocalized and is situated in a region between one Zr and two Si atoms. The corresponding linear spreads amount to 3.1 and 3.8 \AA, respectively. The energy bands being the eigenvalues of the Wannier Hamiltonian are shown in
Fig.~\ref{bands}, from which one can see a good match between the DFT and Wannier projected bands in the vicinity of the Fermi energy. Importantly, by projecting the Bloch states $\psi^{n}_{\bf k}({\bf r})$ onto the resulting WF, one can see that $| \psi^{1}_{\bf k}\rangle \approx  |w^{(A)}_{\bf k}\rangle$ and $|\psi^{2}_{\bf k} \rangle \approx  | w^{(B)}_{\bf k} \rangle$, meaning that each Bloch band is primarily composed of a single WF, and that the overlap between different Bloch bands $\langle \psi^1_{\bf k}|\psi^2_{\bf k} \rangle$ is essentially zero at any {\bf k} point, as depicted by the color of the bands in Fig.~\ref{bands}.

\begin{figure}[b!t]
\includegraphics[width=0.95\linewidth]{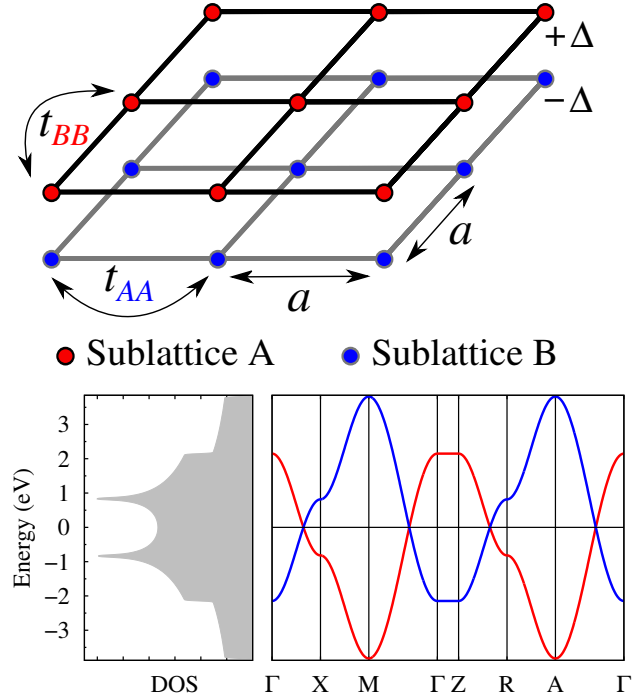}
\caption{(Top) Schematic illustration of the lattice TB model used in this Letter to describe single-particle states in \mbox{ZrSiS}. (Bottom) Band structure and density of states of the corresponding model.}
\label{model}
\end{figure}

The resulting lattice model for \mbox{ZrSiS} is schematically shown in Fig.~\ref{model}. The model is given by two nested square sublattices ($A$ and $B$) in two dimensions.
The corresponding Hamiltonian reads
\begin{align}
H = \sum_{ij \in A} \left(t^{\phantom{\dag}}_{ij}+\Delta\delta^{\phantom{\dag}}_{ij}\right) a_i^{\dag} a^{\phantom{\dag}}_j + \sum_{ij \in B} \left(t^{\phantom{\dag}}_{ij}-\Delta\delta^{\phantom{\dag}}_{ij}\right) a_i^{\dag} a^{\phantom{\dag}}_j,
\label{hamilt}
\end{align}
where $a_i^{\dag}$ ($a_i$) is the creation (annihilation) operator, $\Delta=0.835$ eV is the on-site energy shift originating from the difference between the sublattices, and
$t_{ij}$ are the dominant hoppings parameters, which are restricted to the next-nearest-neighbor interactions within each of
the sublattice: $t_{AA}$ and $t_{BB}$. Remarkably, $t_{AA} \approx -t_{BB} \approx 0.74$ eV, while the intersublattice hopping amplitude is considerably smaller, $t_{AB} \sim 0.1$ eV $\ll t_{AA}$, and, therefore, it is not considered here, which significantly simplifies
the model. Although some changes in the band structure shown in Fig.~\ref{model} are noticeable (cf. Fig.~\ref{bands}), the main characteristics of the spectrum remain unchanged. It is important
to emphasize that the Hamiltonian given above is diagonal in the sublattice basis. In {\bf k} space, the diagonal elements have the form
\begin{equation}
\varepsilon_{\pm}({\bf k}) = \pm 2t_{AA}[\mathrm{cos}k_xa + \mathrm{cos}k_ya] \mp \Delta,
\label{eq:dispersion}
\end{equation}
where $a=3.55$ \AA~is the lattice constant in \mbox{ZrSiS}. Importantly, the absence of off-diagonal elements means, particularly, that the given model would not reproduce a nontrivial $\pi$ Berry phase observed experimentally \cite{Wang,Ali,Singha,Pezzini}. In order to restore the topologically nontrivial behavior, one would need to include the $t_{AB}$ hopping parameter in the model~\cite{Young}. However, in the context of strong correlations, the small magnitude of $t_{AB}$ allows us to neglect it, at least, as the first approximation, when considering many-body phenomena of interest. 

As it was shown in Refs.~\onlinecite{Keldysh,Kozlov}, a semimetal with congruent electron and hole Fermi surfaces exhibits a strong instability of the ground state with respect to the electron-hole coupling, which is  mediated by the Coulomb interaction. Formally speaking, this instability is similar to formation of Cooper pairs in superconductors in the sense that electron-hole symmetry is playing the same role as time-reversal symmetry for the superconductors. It is not easy, however, to find a system with high enough degree of electron-hole symmetry. Graphene~\cite{Graphene} is the most well-known example; however, a vanishing density of states near the neutrality point suppresses essentially the potential instabilities \cite{Sabio,Ulybyshev}. Bilayer graphene is another example for the realization of the excitonic instabilities, widely discussed in the literature \cite{Levitov,Mayorov,Bitan,Maher,McDonald}.

Within our model for ZrSiS, the dispersion of electrons~[Eq.~\eqref{eq:dispersion}] is diagonal in the sublattice $\{A,B\}$ space, and it exhibits a high degree of electron-hole symmetry $\varepsilon_{A}(\kv)=-\varepsilon_{B}(\kv)$.
Furthermore, the nodal line band topology ensures a finite density of states at the Fermi energy, which makes \mbox{ZrSiS} a natural candidate for the excitonic insulator or, at least, for a system in the vicinity of a quantum phase transition to the excitonic insulator phase (``virtual'' excitonic insulator). Therefore, it is instructive to consider electron-electron interactions explicitly, which is done in this Letter at the level of the two-orbital Hubbard model.

{\it Coulomb interaction}.---Before considering many-body effects in \mbox{ZrSiS}, we estimate the strength of the Coulomb interactions in this material. As we are aiming at solving the two-orbital Hubbard model, we are interested in the partially screened Coulomb interaction $U=(1-VP)^{-1}V$, where $V_{ij}=\langle ij | |{\bf r-r'}|^{-1} |ij \rangle$ is the bare interaction, and $P$ is the static polarizability calculated here within the constrained random phase approximation \cite{cRPA,noteRPA}. The resulting on-site and intersite Coulomb-interaction parameters estimated to $U_{AA}=2.00$, $U_{BB}=1.22$, and $U_{AB}=0.97$ eV, respectively, which are determined in the WF basis defined above. As expected, the on-site interactions are different for the two sublattices, which is attributed to different orbital localization. In contrast to typical transition metal compounds, where intersite Coulomb interaction within the $d$ shell is essentially screened by conduction electrons \cite{Solovyev}, intersite coupling in ZrSiS is not more than two times smaller than the on-site interaction, i.e., $U_{ii}/U_{ij} \lesssim 2$, typical to $p$-electron systems \cite{Wehling,Mazurenko,Rudenko}. The presence of a non-negligible interaction between the sublattices will result in a Coulomb mixing of the relevant low-energy states in \mbox{ZrSiS}.

Consideration of Coulomb interactions leads to the following action of the two-orbital Hubbard model written in momentum space
\begin{align}
\label{eq:action}
{\cal S} = &-\sum_{{k},\alpha} a^{*}_{{k}\,\alpha} \left[i\nu+\mu^{\phantom{*}}_{\alpha} - \varepsilon^{\phantom{*}}_{\alpha}(\kv)\right]
\, a^{\phantom{*}}_{{k}\,\alpha} \\
&+\sum\limits_{{q},\alpha} U_{\alpha\alpha}\, n_{{q}\,\alpha\uparrow}\, n_{-{q}\,\alpha\downarrow}
+\frac12\sum\limits_{{q},\alpha,\beta} U_{\alpha\beta}\,n_{{q}\,\alpha} \,n_{-{q}\,\beta}. \notag
\end{align}
Here, variables $a^{*}_{{k}\,\alpha}$ ($a^{\phantom{*}}_{{k}\,\alpha}$) correspond to creation (annihilation) of an electron with momentum $\kv$ and fermionic Matsubara frequency $\nu$ on a sublattice $\alpha=\{A, B\}$. $n_{{q}\,\alpha}=\sum_{{k}}a^{*}_{{k}\,\alpha}a^{\phantom{*}}_{{k+q}\,\alpha}$ is an electronic density for the momentum $\qv$ and bosonic Matsubara frequency $\omega$. For simplicity, we use the combined notations for momentum and frequency ${k}=\{\kv,\nu\}$ and ${q}=\{\qv,\omega\}$ for electronic and bosonic variables, respectively. Then, the bare Green's function of the model is diagonal in the sublattice $\alpha=\{A,B\}$ space and reads
${\cal G}_{\alpha}({k}) = [i\nu+\mu_{\alpha}-\varepsilon_{\alpha}(\kv)]^{-1}$.

{\it Formation of a pseudogap}.---Similar to the case of superconductivity, the electron-hole instability \cite{Keldysh,Kozlov} takes place at decreasing temperatures. At the critical temperature $T_{c}$, the system may undergo the second-order phase transition toward the symmetry broken phase, characterized by anomalous averages; for the case of superconductivity, they describe Cooper pairing, while in the case under consideration, they couple electron and hole states. The ordered phase is characterized by the formation of the bound electron-hole pairs, which leads to a (pseudo)gap opening in the electronic spectrum. From the symmetry point of view, this works like an effective spin-orbit coupling, yet without the relativistic smallness parameter (i.e., can yield a much larger gap) and being strongly temperature dependent. Above the critical temperature, a pseudogap can arise due to fluctuation effects, similar to the antiferromagnetic pseudogap in the quantum two-dimensional antiferromagnet at finite temperatures~\cite{IrKats}. Unlike the case of a superconductor, where the fluctuation effects are small due to the long-range character of the Cooper pairing, 
the situation may be different for the case of excitonic pairing since there are no reasons to {\it a priori} assume a very large exciton radius. 

To capture the relevant fluctuation effects in ZrSiS, let us introduce the electronic self-energy $\Sigma_{\alpha}({k})$
that takes correlation effects in the particle-hole (excitonic) channel into account. In order to describe the main contribution that leads to the instability, the self-energy is taken in the following form
\begin{align}
\includegraphics[width=0.90\linewidth]{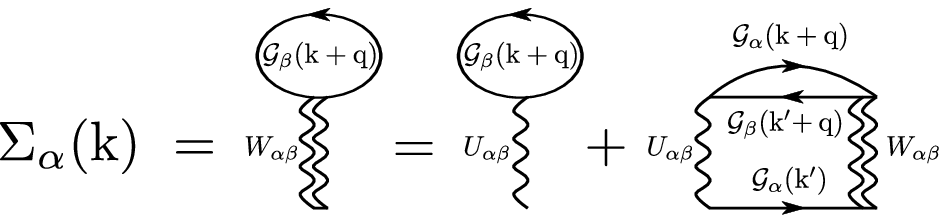},
\end{align}
or
\begin{align}
\Sigma_{\alpha}({k}) = \sum_{{q},\,\beta} W_{\alpha\beta}({q})\, {\cal G}_{\beta}({k+q}),
\label{eq:selfen}
\end{align}
where the renormalized interaction
\begin{align}
W_{\alpha\beta}({q}) = U_{\alpha\beta}\,\left[1 - X^{0}_{\alpha\beta}({q})U_{\alpha\beta}\right]^{-1}
\label{eq:Wren}
\end{align}
contains the two-particle ladder diagrams that couple electrons of different sublattices
\begin{align}
X^{0}_{\alpha\beta}({q}) = \sum_{{k}'} {\cal G}_{\alpha}({k}')\, {\cal G}_{\beta}({k}'+{q})
\label{eq:Xlad}
\end{align}
via the Coulomb interaction.
The calculation of the self-energy [Eqs.~(\ref{eq:selfen})$-$(\ref{eq:Xlad})] is performed on a $128\times128$ $\kv$ mesh with 512 Matsubara frequencies using the fast Fourier transform scheme, similar to Ref.~\onlinecite{biFLEX}.
Then, the renormalized Green's function is given by the usual Dyson equation
\begin{align}
G^{-1}_{\alpha}({k}) = {\cal G}^{-1}_{\alpha}({k}) - \Sigma^{\phantom{1}}_{\alpha}({k}).
\label{eq:Gfull}
\end{align}
Finally, we perform a Pad\'{e} continuation~\cite{Vidb77} of the renormalized Green's function $G_{\alpha}(\kv,\nu)$ from the Matsubara axis to the real frequencies $G(\kv,E)$ and calculate the spectral function $A({\bf k},E)=-\,(1/\pi) \,\mathrm{Im}\,G({\bf k},E)$ for different temperatures. At high enough temperatures, no significant changes of the spectral function are observed.

\begin{figure}[t!]
\includegraphics[width=1\linewidth]{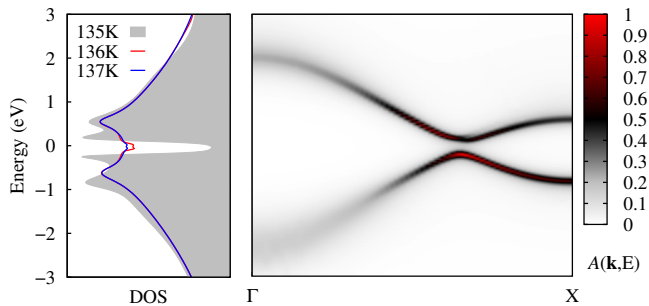}
\caption{Renormalized spectral function $A({\bf k},E)=(1/\pi) \mathrm{Im} G({\bf k},E)$ and density of states of \mbox{ZrSiS} calculated near the transition temperature $T_c=135$ K at which a pseudogap is formed. }
\label{Ak}
\end{figure}

In Fig.~\ref{Ak}, we show the resulting spectral function, as well as the density of states, calculated in the vicinity of the phase transition. One can see that at $T_{c}\approx135$ K the band crossing between the $\Gamma$ and $X$ points disappears, accompanied by a pronounced dip in the density of states. Therefore, one can see that there is a phase transition induced by many-body effects. Indeed, similar to~\cite{IrKats}, the renormalized interaction \eqref{eq:Wren} that couples electrons from two different sublattices $W_{AB}({q\to0})$ depends on the temperature and diverges in the region close to the phase transition. This dependence follows from the logarithmic divergence of $X^0(q\to0) \propto$ $-N(0)\,\mathrm{ln}(\Delta/T)$ for zero doping, where $N(0)$ is the density of states at zero Fermi energy, and $\Delta$ is a cutoff of the order of bandwidth \cite{Keldysh,Kozlov}. As a result, $W_{AB}({q\to0}) \approx 1/\mathrm{ln}(T/T_c)$. 
In Ref.~\onlinecite{Pezzini}, the effect of a band renormalization was observed in strong magnetic fields. According to the experiment, the effective mass in ZrSiS depends not only on $T$, but also on $B$, which is the magnitude of the magnetic field. It can be shown that the magnetic field cuts off the logarithmic divergence responsible for the excitonic instability at energies on the order of the cyclotron frequency, $\hbar \omega_c \sim B$. It means that in the expression for $X^0$ given above, $T$ is to be replaced by $\mathrm{max}\{T,\hbar \omega_c\}$. Similar behavior was recently theoretically predicted and experimentally observed for the Fermi velocity renormalization in graphene \cite{Sonntag}.

\begin{figure}[t!]
\includegraphics[width=1\linewidth]{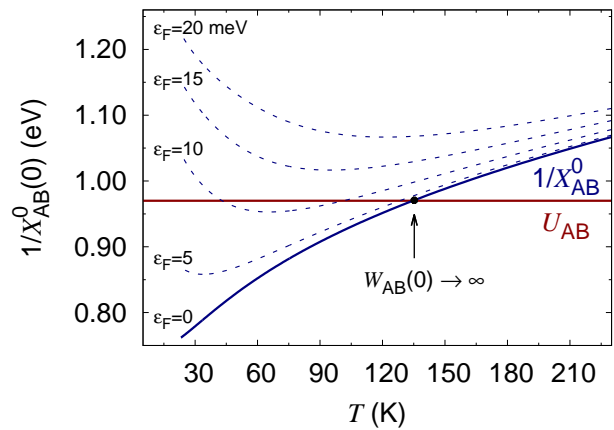}
\caption{Inverse intersublattice susceptibility $1/X^{0}_{AB}(q\to0)$ calculated as a function of temperature for different Fermi energies corresponding to different values of hole doping. Intrasublattice Coulomb interaction ($U_{AB}$) estimated within RPA is shown for reference.  }
\label{chi_T}
\end{figure}

The calculated temperature dependence of $1/X^{0}_{AB}(q\to0)$ relative to the intrasublattice Coulomb interaction $U_{AB}$ is shown in Fig.~\ref{chi_T}. The condition $U^{\phantom{1}}_{AB}\,X^{0}_{AB}(q\to0)=1$ in Eq.~(\ref{eq:Wren}) ensures a large contribution to the self-energy~\eqref{eq:selfen}. At zero doping it results in the formation of the pseudogap in a narrow region of temperatures close to $T_{c}\approx135$ K, as shown in Fig.~\ref{Ak}. As one can also see from Fig.~\ref{chi_T}, the critical temperature $T_c$ strongly depends on doping, which breaks the particle-hole symmetry and suppresses the instability. Even a small shift of the Fermi energy by $\varepsilon_F=10$ meV results in a considerable decrease of $T_c$, whereas at higher values of $\varepsilon_F$, the phase transition becomes fully suppressed. This behavior is similar to suppression of singlet superconductivity by the magnetic field.

It is worth noting that the resulting phase diagram is very sensitive to detailed values of the parameters and especially to the intersublattice coupling parameter $U_{AB}$. Given that long-wavelength correlations are not considered explicitly in the model (\ref{eq:action}), one may expect that the properly defined parameter $U_{AB}$ can be further reduced by screening due to nonlocal interactions neglected in our formalism \cite{Schuler}. In turn, the reduction of $U_{AB}$ can lead to a considerable lowering of the transition temperature $T_c$, what immediately follows from Fig.~\ref{chi_T}. However, we note that the transition to the phase of excitonic pairing cannot be fully suppressed as long as particle-hole symmetry is strictly preserved. Probably only experiment can decide whether this condition is achievable in practice, and whether ZrSiS becomes a truly excitonic insulator at low enough temperatures or just close enough to the instability. The existence of an excitonic pseudogap cannot be conclusively supported by the reported angle-resolved photoemission spectroscopy data (e.g., Refs.~\onlinecite{Neupane,Schoop}), which might be related to the limited resolution, not sufficiently low temperatures, or external factors breaking the electron-hole symmetry (e.g., doping). Nevertheless, our results appear sufficient to explain qualitatively a strong renormalization of the effective mass observed in Ref.~\onlinecite{Pezzini}. Further experiments, such as infrared optical measurements or scanning tunneling microscopy, may be needed to provide more consistent information on the many-body effects in ZrSiS.

{\it Conclusion}.---To conclude, we have demonstrated that ZrSiS is a system with moderately strong electron-electron interactions and a very high degree of electron-hole symmetry near the Fermi energy, which makes it close to excitonic instability. We predict a pseudogap formation in its electronic spectrum at low enough temperatures. Therefore, ZrSiS is a good example of the system where both nontrivial topology and correlation effects are important, which should motivate further experimental and theoretical studies. \\

\begin{acknowledgments}
A.N.R. and E.A.S. acknowledge support from the Russian Science Foundation, Grant No.~17-72-20041.
The work of M.I.K. was supported by NWO via Spinoza Prize and by ERC Advanced Grant No.~338957 FEMTO/NANO. A.I.L. acknowledges support from the  Cluster of Excellence ``The Hamburg Centre for Ultrafast Imaging (CUI)''
of the German Science Foundation (DFG). Support from the Stichting voor Fundamenteel
Onderzoek der Materie (FOM), which is financially supported by the Nederlandse Organisatie voor Wetenschappelijk Onderzoek (NWO) is also acknowledged.
\end{acknowledgments}


\begin{thebibliography}{99}
\bibitem{wannier_code}
For this purpose, we used the {\sc wannier90} code \cite{wannier90}.
\bibitem{noteRPA}
In the calculation of the RPA polarizability matrix $P$, we used a sufficiently large number of unoccupied DFT bands within the energy window of $\sim$50 eV, which ensures numerical convergence of the Coulomb-interaction parameters.
\end{thebibliography}
\end{document}